# NESTML: a modeling language for spiking neurons

Dimitri Plotnikov[1], Bernhard Rumpe[1], Inga Blundell[2], Tammo Ippen[2,3], Jochen Martin Eppler[4] and Abigail Morrison[2,4,5]

**Abstract:**

Biological nervous systems exhibit astonishing complexity. Neuroscientists aim to capture this complexity by modeling and simulation of biological processes. Often very complex models are necessary to depict the processes, which makes it difficult to create these models. Powerful tools are thus necessary, which enable neuroscientists to express models in a comprehensive and concise way and generate efficient code for digital simulations. Several modeling languages for computational neuroscience have been proposed [Gl10, Ra11]. However, as these languages seek simulator independence they typically only support a subset of the features desired by the modeler. In this article, we present the modular and extensible domain specific language NESTML, which provides neuroscience domain concepts as first-class language constructs and supports domain experts in creating neuron models for the neural simulation tool NEST. NESTML and a set of example models are publically available on GitHub.

## 1 Introduction

Classical neuroscience investigates the biophysical processes behind single neuron behavior and higher brain function. The first experimental studies of the nervous system were conducted already hundreds of years ago [Sh91], but observing single-cell activity in a cell culture or slice (*in vitro*) or in an intact brain (*in vivo*) is a technically challenging task. It was thus not before the beginning of the last century that details about the structure and function of the building blocks of the brain became known.

In the early 40s of the last century, McCulloch and Pitts [MP43] explored the idea of using simple threshold elements to mimic the behavior of interconnected nerve cells. However, it soon turned out that these artificially built circuits were too simple and limited to study the principles at work in living brains.

[1] RWTH Aachen University, Chair of Software Engineering, Jülich Aachen Research Alliance (JARA), Ahornstraße 55, 52074 Aachen, Germany

[2] Forschungszentrum Jülich, Institute of Neuroscience and Medicine (INM-6), Institute for Advanced Simulation (IAS-6), JARA BRAIN Institute I, 52025 Jülich, Germany

[3] Norwegian University of Life Sciences, Dept. of Mathematical Sciences and Technology, 1432 Ås, Norway

[4] Forschungszentrum Jülich, Simulation Lab Neuroscience, Bernstein Facility for Simulation and Database Technology, Institute for Advanced Simulation, JARA, 52025 Jülich, Germany

[5] Ruhr-University Bochum, Faculty of Psychology, Institute of Cognitive Neuroscience, 44801 Bochum, Germany

The field of neural networks consequently split: the descendants of the early neural networks are still used under the term *artificial neural networks* (ANN) to solve learning and classification tasks in engineering applications. Biologically more plausible models of neural circuits are nowadays known as *spiking* or *biological neural networks*.

## 1.1    Neural modeling and simulation

Computational neuroscience builds models for nerve cells (*neurons*) and their connections (*synapses*) that capture certain aspects of their anatomy and physiology. Depending on the study, different aspects are important (Section 2). As theoreticians prefer to use the simplest model that still exhibits the behavior they are interested in, a multitude of different models was published. The level of detail ranges from *compartmental models* that include many biophysical details to reduced *point neuron models* that describe the basic quantities or the cell by a small set of differential equations (Section 2). The simulation of networks of such model neurons (i.e. the propagation of the underlying equations in time) allows to execute *in silico* experiments to test hypotheses in a stable and controllable environment.

As the simulation of different classes of neurons requires different technical infrastructure (e.g. for the storage of connections or the communication between elements), different simulators have been developed. Each of them is specialized on a specific part of the spectrum of modeling tasks. This makes it hard to develop new neuron and synapse models in a general way and even harder to compare and verify findings across simulators, since models must be re-implemented for every simulator [Cr12].

To ease model-sharing and improve reproducibility in the field, several modeling languages were conceived (Section 3). They usually consist of the language itself and tools to generate a model implementation from a model specification. As the majority of the languages are simulator agnostic, they cannot take advantage of the convenience functions of a given simulator. This often results in models with lower performance or accuracy compared to a hand-written version of the same model.

## 1.2    The neural simulation tool NEST

NEST [GD07] is a simulator for large networks of spiking point neurons available as open source software (www.nest-simulator.org). Using hybrid parallelization it runs on all machines from laptops to the world's largest supercomputers [He12, Ku14]. Over 450 published studies used NEST and 360 users are currently subscribed to the mailing list. Due to its reliability and popularity, NEST was selected as simulator for brain-scale networks of simplified neurons in EU's Flagship *Human Brain Project* (http://humanbrainproject.eu).

At the outset of this study, NEST contained 36 neuron models, each of which implemented by hand as a C++ class using NEST's model API and embedded into NEST's infrastructure. Developing new models requires expert knowledge of the neuroscience context, as well as of C++ and NEST's internals. Changes to NEST's infrastructure or API often require changes to all models, which impairs the maintainability of NEST.

The C++ classes mix the model description (i.e. the equations and algorithms governing the dynamic behavior) with the model implementation, which impairs model comprehension. An example are the linear models in NEST, which use an exact solution for the differential equation rather than one obtained by a general solver [RD99]. This hides the actual equations deep in the model code.

Due to the lack of modularity in the C++ model code, new neuron models are mostly created by *copy&paste* from existing models. The fact that this task is often carried out by neuroscientists who are not experts in programming leads to redundancy, suboptimal performance, improper documentation and reduced maintainability. Preliminary investigations show cases where two models share more than 90% of their implementation.

### 1.3   The NEST modeling language NESTML

NESTML is a domain specific language that supports the specification of neuron models in a precise and concise syntax, which is familiar to the domain experts. Model equations can either be given as a simple string of mathematical notation or as an algorithm written in the built-in procedural language. The equations are analyzed by NESTML to compute an exact solution if possible or use an appropriate numeric solver otherwise (Section 4).

The simplicity of the explicit syntax of NESTML guarantees good comprehensibility and a clear separation between the model specification and its implementation. A code generator creates optimized model code alongside auxiliary code to load the model dynamically into NEST (Section 5).

First class modularization concepts in the language simplify the reuse of neuron definitions and parts thereof. This feature fosters the re-use of well tested components in models instead of re-implementing them. Models expressed in other languages can be compiled to NESTML by the code generation tools of the language.

Being built on top of the language workbench MontiCore [KRV07, KRV08], all tools belonging to NESTML are generated from a language grammar. This allows us to conveniently update the language itself to new modeling requirements, and the code generator to changes in the NEST infrastructure or API.

## 2   Modeling spiking neurons

As with all body cells, neurons are also confined by a membrane. *Channels* embedded into the membrane selectively allow certain types of ions to pass, active transporter molecules move ions in and out of the cell. These mechanisms maintain up a gradient of charges, resulting in an electrical potential across the membrane.

An incoming signal (*action potential*, *spike*) leads to a short excursion of the membrane potential. The direction of the excursion depends on the type of the sending (*presynaptic*) neuron, which can be either *excitatory* (positive excursion) or *inhibitory* (negative excursion). If the input is strong enough or several inputs occur simultaneously, the membrane potential eventually reaches a *threshold* and the neuron fires a spike itself. Spikes are transmitted via the synapses to receiving (*postsynaptic*) neurons, where the spike again leads to a change of the membrane potential. After emitting a spike, a neuron is inactive for a certain time, called its *refractory period* [Ni01].

The work of Lapicque [La07] and later Hodgkin and Huxley [HH52] paved the way for creating models of neurons with biologically realistic parameters. For the membrane potential, they define an equivalent electrical circuit, in which the membrane itself is represented by a capacitor, ion channels by resistors and external inputs by an additional current.

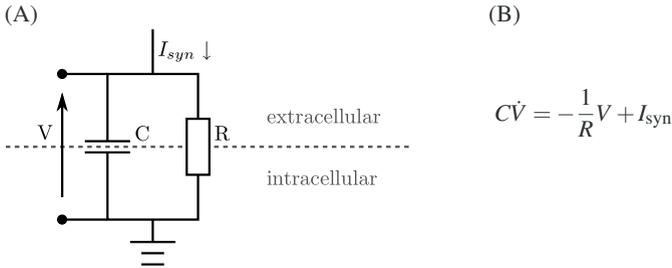

(A)     (B)

$$C\dot{V} = -\frac{1}{R}V + I_{\text{syn}}$$

Figure 1: Electrical circuit corresponding to single compartment of a neuron. (A) circuit diagram. (B) differential equation for the membrane potential $V$ given capacitance $C$, resistance $R$ and external input current $I_{\text{syn}}$

.

A common approach to modeling neurons is to divide the 3D reconstruction of a real neuron into compartments and use one Hodgkin and Huxley circuit for each compartment. The compartments are coupled using the formalism of cable theory. In case the morphology of such a *multi-compartment model* only consists of a single compartment, it is called a *point neuron model*.

The basic differential equation shown in Figure 1 only characterizes the subthreshold dynamics of the neuron. Checks for threshold crossings, spike generation and refractoriness are usually added in an algorithmic fashion using conditionals and wait cycles. Spiking input enters the equation in form of a summed current $I_{\text{syn}}$. To obtain its value, the time of each incoming spike is convolved with a kernel that represents the excursion of the membrane potential (*post-synaptic potential*). Frequently used functions for this kernel are $\alpha$-shapes with varying time constants, exponentially decaying functions or delta pulses. Multiple inputs are lumped together into $I_{\text{syn}}$ before propagation of the equation to the next time step. This approach is commonly referred to as *current-based* modeling.

Another way to model the influence of external input is the *conductance-based* approach. In contrast to integrating the inputs into a general current variable, the input instead influences the conductance of the membrane in this case. This is generally considered more realistic, but leads to a non-linear differential equation, as changes to the conductance depend on the membrane potential and vice versa. Simulating these models is computationally more demanding than it is for the current-based approach.

Due to their property of integrating incoming spikes and firing one if a threshold is reached, the family of models described above is known under the name *integrate-and-fire neurons*.

## 2.1 Neuron dynamics

Substituting the resistor and the capacitor of the RC circuit shown in Figure 1 by the rise time $\tau_m$ (*membrane time constant*) and the capacitance $C$, we obtain the following equation for the membrane potential $V$ of the standard integrate-and-fire neuron:

$$\frac{d}{dt}V = -\frac{V}{\tau_m} + \frac{1}{C}I \tag{1}$$

The input current $I$ is the sum of the synaptic current and any external input. The $\alpha$-shaped synaptic current as a function of time $t$ for one incoming spike is given by:

$$\iota(t) = \hat{\iota}\frac{e}{\tau_\alpha}te^{\frac{-t}{\tau_\alpha}}. \tag{2}$$

Here $\hat{\iota}$ is the peak value of the incoming spike and $\tau_\alpha$ is the rise time. The inhomogeneous differential equation (2) (for simplicity we assume that $I == \iota$) is rephrased as a homogeneous system of differential equations (or matrix differential equation):

$$\frac{d}{dt}\begin{pmatrix} \frac{d}{dt}\iota + \frac{1}{\tau_\alpha}\iota \\ \iota \\ V \end{pmatrix} = \begin{pmatrix} -\frac{1}{\tau_\alpha} & 0 & 0 \\ 1 & -\frac{1}{\tau_\alpha} & 0 \\ 0 & \frac{1}{C} & -\frac{1}{\tau_m} \end{pmatrix} \cdot \begin{pmatrix} \frac{d}{dt}\iota + \frac{1}{\tau_\alpha}\iota \\ \iota \\ V \end{pmatrix} \tag{3}$$

For a fixed time step $t$ it is now possible to solve the differential equation by calculation of the matrix exponential of the given matrix [RD99]. This way of propagating the model in time is particularly efficient because it consists only of a few multiplications.

Although this calculation is done only once for each linear neuron model in NEST during the implementation of the model, it is tedious and has to be done manually. With NESTML, all necessary factors for the time propagation for any given synaptic current and any linear differential equation can be calculated automatically, which solves one of the major obstacles for developing new neuron models in NEST.

## 3 Related Work

Various modeling languages for neurons and neural networks exist, each of which focusing on different aspects of neural modeling. Here, we describe the representative examples NineML and NeuroML in detail.

## 3.1   NineML

The Network Interchange for Neuroscience Modeling Language (NineML, [Ra11, Go11]) provides an unambiguous description of spiking neural networks for model sharing and re-use. NineML defines a common object model that describes the different elements of a model in a neuronal network. This object model corresponds to its abstract syntax, while XML is used as its concrete syntax.

NineML consists of two semantic layers: the abstract layer describes the core concepts of a model alongside its mathematical description, parameter and state variables and state update rules. The user layer allows the description of state or parameter variables and definition of initial or default values and units. Objects defined in the user layer can be re-used in different models, while model re-use in the abstract layer is not supported.

In the abstract layer each network element is represented by a `ComponentClass` composed of a `Dynamics`-block and a set of `Interfaces`. The `Dynamics` contain the internal model dynamics, e.g. state variables and update rules. The `Interfaces` contain the parameters that can be set from the user layer and ports for the communication with other network elements. The advantage of the `ComponentClass` is that it supports any kind of network element instead of just complete neuron or synapse models. The drawback is that the exact kind of model modeled by the `ComponentClass` is unknown. It could be a neuron, a synapse or an ion channel and the relation to domain concepts is hidden from the user.

To make NineML descriptions simulator agnostic, they only provide differential equations to describe the dynamics of a model. As the system itself chooses the solver for the dynamics, this might lead to the generation of unnecessarily complex and inefficient code for a specific simulator or to an inaccurate solution of the model equations. Expressing neuron dynamics as a finite-state automaton with regimes and transitions as in NineML works well to visualize them. However, for developing new neuron models and expressing complex relationships between states a procedural definition of the dynamics is more intuitive.

## 3.2   NeuroML

The model description language NeuroML [Gl10] is a description language for biophysically detailed neuron and neural network models and enables interoperability across multiple simulators. Neuron models in NeuroML can have complex morphologies, voltage- and ligand-gated conductances, and synaptic mechanisms. Network models contain the 3D positions of cells and synapses in the network.

NeuroML is optimized for complex compartmental models, but also supports simple point neurons like the leaky integrate-and-fire model (Section 2). However, more advanced types of point neuron models such as the exponential integrate-and-fire neuron [BG05] or the Izhikevich model [Iz03] are not fully supported yet. The language itself is split in three levels, each of which is responsible for describing a different scale of biological detail:

**Level 1** describes the morphology of a neuron model using the sub-language MorphML. This contains the number and 3D position of compartments and their size and shape. Additionally, it provides mechanisms to store metadata.

**Level 2** uses ChannelML to describe voltage-gated membrane conductances together with static and plastic synaptic conductance processes. It also extends level 1 descriptions by specifying the location and density of membrane conductances in the cell model.

**Level 3** describes neural networks with 3D locations of individual neurons, synaptic connections between neurons (in projections) and external inputs via NetworkML.

NeuroML can define neuron models by using predefined elements for segments, channel mechanisms or synapse mechanisms. This results in compact and clear definitions of models by outsourcing and reusing mechanism definitions. On the other hand, the limited set of possible language elements reduces the expressiveness of NeuroML to models for which corresponding elements exist. Defining new mechanisms requires changes to the language definition itself.

### 3.3 XML as carrier language

Most of the established modeling languages use XML [Ye04] as their concrete representation, because an ecosystem of tools already exists and no additional lexers and parsers have to be developed to check syntactic correctness. However, this approach has two disadvantages: first, the verbosity of XML makes writing and reading models difficult for modelers [Ch01] and sophisticated tools are required for creating, visualizing and understanding more complex models. Second, the model descriptions have to be processed separately to ensure semantic correctness. An example for this is NineML's `MathInline` statement, which requires custom parsers to check the contained mathematical expressions for correctness. Listing 1 illustrates these two problems of XML using an excerpt of a NineML file.

```
 1  ...
 2  <Dynamics>
 3    <StateVariable name="V" dimension="voltage" />
 4    <StateVariable name="U" dimension="voltage␣per␣time" />
 5    <Alias name="rv" dimension="none">
 6      <MathInline>V*U</MathInline>
 7    </Alias>
 8    <Regime name="subthresholdRegime">
 9      <TimeDerivative variable="U">
10        <MathInline>a*(b*V - U)</MathInline>
11      </TimeDerivative>
12      <TimeDerivative variable="V">
13        <MathInline>0.04*V*V + 5*V + 140.0 - U + iSyn</MathInline>
14      </TimeDerivative>
15    </Regime>
16  </Dynamics>
17  ...
```

Listing 1: Excerpt from a NineML file. To declare the simple mathematical expression $rv = V * U$, three lines of code are required (cf. lines 5-7). The `MathInline` element in line 13 contains only a string that cannot be checked for syntactic or semantic correctness with existing XML tools.

## 3.4   Simulators

Before the existence of general model description languages, simulators already had their own languages for specifying models. In the case of NEST (Section 1.2) this language so far is just plain C++ and the features provided by the simulator API. The remainder of this section introduces two other approaches for the definition of neuron and network models for completeness.

Brian [St14] is a simulator for spiking neural networks written entirely in Python. It uses code generation based on SymPy, NumPy and Cython to obtain reasonable performance, but lacks the facilities for running distributed simulations. Neuron models are defined by specifying the differential equations written in a text-based mathematical notation. However, as these definitions are ordinary Python strings, checking context conditions and semantically analyzing them is difficult. Unless own extensions to Brian are provided, it is up to the simulator to chose a solver method for the equations, which can have negative effects on accuracy or efficiency. Brian is mainly used for small-scale and exploratory simulations on laptops and workstations.

NEURON [HC97] is a simulator mainly for compartmental neuron models with biophysical properties. Neuron and synapse models can be defined with a set of graphical tools or using the custom programming language HOC. NEURON's focus is not on large-scale modeling, but on the simulation of very detailed neuron models on large computer clusters and supercomputers. In principle, it also supports simulations of large networks of simple neuron models, but falls behind the performance and memory footprint of simulators that are aimed specifically at these simulations.

# 4   Modeling spiking neurons with NESTML

NESTML consists of three modular and separately usable sub-languages, a symbol table and context conditions. These languages together form the NESTML domain specific language (DSL).

**Procedural DSL (PL)** defines the imperative logic of the model. PL also provides a library with methods for emitting messages, logging and working with buffer objects.

**Units DSL (UL)** enables defining and checking variables with physical units like Volt (V) and Ampere (A). UL also supports common magnitudes like mili (m) and pico (p).

**Differential Equation DSL (DL)** provides the possibility to define differential equations in the form of a string of math notation and analyze these equations.

NESTML separates model definition from simulator specific code and thereby allows the user to concentrate on the development of models instead of implementation details. Automatic analysis of differential equations simplifies the formulation of new models by outsourcing the task of finding an accurate solution to NESTML's infrastructure. This section introduces NESTML with the example of a simple integrate-and-fire neuron [RD99].

## 4.1    Basic design and definitions

The general syntax of NESTML is inspired by that of Python, which is widely known to researchers in the computational neuroscience community [Mu09, Da13]. This lowers the entry barrier for new users and improves comprehensibility of models. NESTML supports common data types like *integer*, *real* and *string* as well as physical data types with units provided by the PL. Variables are defined by stating the name followed by a type or unit.

```
① neuron iaf_neuron:

②   state:
        y0, y1, y2, y3, V_m mV [V_m >= -99.0]
        # Membrane potential
        alias V_rel mV = V_m + E_L
     end

③   function set_V_rel(v mV):
        y3 = v - E_L
     end

④   parameter:
        # Capacity of the membrane.
        C_m     pF = 250 [C_m > 0]
     end

⑤   internal:
        h    ms  = resolution()
        P11 real = exp(-h / tau_syn)
        ...
        P32 real = 1 / C_m * (P33 - P11)
                   / (-1/tau_m - -1/tau_syn)
     end
```

```
⑥   input:
        spikeBuffer    <- inhibitory
                          excitatory spike
        currentBuffer <- current
     end

⑦   output: spike

⑧   dynamics timestep(t ms):
        if r == 0: # not refractory
          V_m = P30 * (y0 + I_e) + P31 *
              y1 + P32 * y2 + P33 * V_m
        else:
          r = r - 1
        end
        # alpha shape PSCs
        V_m = P21 * y1 + P22 * y2
        y1 = y1 * P11
        y0 = currentBuffer.getSum(t);
     end

  end
```

(Excerpt from the explicit ODE solution)

Figure 2: Excerpt from the integrate-and-fire neuron expressed in NESTML. See https://github.com/nest/nestml for the complete neuron model description.

A neuron in NESTML is declared by the keyword `neuron` and a name (① in Figure 2). The name can be used to reference the model from other NESTML models. Each neuron is composed of blocks with definitions of *state* and *parameter* variables, *inputs* and *outputs*. A *dynamics* function is responsible for the behavior of the neuron when the model is simulated. All blocks in NESTML start with a colon and end with the keyword `end` state ② contains the variables of the dynamic state of the neuron. An example for a state variable is the membrane potential of a neuron (`V_m`). An `alias` variable describes the dependency between variables using an expression (`V_rel`). For setting a value on an alias a setter function is required (`set_V_rel`), as the defining expression cannot be inverted automatically for the general case. Plausibility constraints can be added in square brackets after the variable definition (`V_m >= -99.0`). These are useful for debugging and during the development phase of the model and can be removed in the production version for better performance.

parameter ④ contains attributes that do not change over time, but may vary among neuron instances. Examples are the length of the refractory period or the membrane capacitance (`C_m`). To ensure that values are in a sensible range, it is possible to define guards which are evaluated every time a parameter is changed by the user. The syntax is the same as for the plausibility constraints in the `state` block.

`internal` ⑤  contains values that depend on the parameters, but can be precalculated once or auxiliary variables needed for the implementation. In Figure 2 for example, the propagator matrix (i.e. the solution of the model equation) is defined in this block.

`input` ⑥  Several named inputs can be declared using the name of the buffer that should receive the specified input during simulation. The input type can specified as `spike` or `current`. A spike input can further be `inhibitory`, `excitatory` or both. Depending on the sign of the input, incoming spikes are routed to the corresponding sub-buffer. If no such modifier is given the buffer receives all spikes.

`output` ⑦  Each neuron in NEST can just send one type of event during simulation. NESTML supports `spike` or `current` output, which is specified after the keyword `output`.

Functions allow the convenient reuse of code (e.g. ③ in Figure 2). Their definition starts with the keyword `function` followed by the function name and a list of zero or more function parameters in parentheses. Just like declaring a variable, a parameter is declared by first stating its name and then its type. Multiple parameters are separated by a comma. The parameter list is followed by an optional return type.

The definition of the dynamics of a neuron is similar to that of a function (e.g. ⑧ in Figure 2). It starts with the keyword `dynamics` followed by the type of the dynamics. Depending on the type, the function is called once per update step (`timestep`) or just once per minimum delay interval in the simulated network (`minDelay`). A list of parameters can be defined in parentheses.

## 4.2  Modularity and component concept

In order to reuse parts of a model they must be defined in a block starting with the keyword `component` and a name (② in Figure 3). The component is then `imported` into a neuron (see ①) and made available using the keyword `use` and optionally giving a convenient name (see ③). Functions and variables from the component can be referenced using the dot-notation (see ④).

```
❶  import PSPHelpers

    neuron iaf_neuron:

❸    use PSPHelpers as PSP

      dynamics timestep(t ms):
❹      PSP.computePSPStep(t)
        # alpha shape PSCs
        y2 = P21 * y1 + P22 * y2
        y1 = y1 * P11
      end

      ...

    end
```

```
❷  component PSPHelpers:
      state:
        - y0, y1, y2, V_m mV [V_m >= 0]
        alias V_rel mV = y3 + E_L
      end

      function computePSPStep(t ms):
        if r == 0: # not refractory
          y3 = P30 * (y0 + I_e) + P31 *
               y1 + P32 * y2 + P33 * y3
        else:
          r = r - 1
        end

      end
      ...
    end
```

Figure 3: An example for a neuron that reuses a function from a component. Left panel: the code of the referencing neuron; right panel: the code of the component.

This concludes the description of the imperative approach, where the solution of the underlying differential equation is described completely and explicitly in the blocks `internal` and `dynamics`. This approach maps directly to the current implementation of models in NEST. In addition, NESTML provides a declarative approach that is more intuitive, because it is closer to the mathematical description of neuron models common in computational neuroscience.

## 4.3  Declarative model definition

One of the main difficulties in writing models for NEST is writing the code for solving the equations, as this requires advanced knowledge of mathematics and numerics. In the declarative approach, differential equations are directly expressed as a string in mathematical notation under an `ODE` block. Figure 4 shows the declaration of an equation for the current (see ③) and the differential equation for the membrane potential `V_m` (see ④). As this is also the way how models are presented in publications, this syntax makes it easy to re-implement published models in NESTML.

```
neuron iaf_neuron:
  internal:
    h    ms  = resolution()
    P11 real = exp(-h / tau_syn)
    ...
    P32 real = 1 / C_m * (P33 - P11)
               / (-1/tau_m - -1/tau_syn)
  end

  dynamics timestep(t ms):
    if r == 0: # not refractory
①    V_m = P30 * (y0 + I_e) + P31 *
           z1 + P32 * y2 + P33 * y3
    else:
      r = r - 1
    end
    # alpha shape PSCs
②  V_m = P21 * y1 + P22 * V_m
    y1 = y1 * P11
  end

end
```

```
neuron iaf_neuron_ode:
  internal:
    h    ms  = resolution()
  end

  dynamics timestep(t ms):
    if r == 0: # not refractory
      ODE:
③      I_shape == w * (E/tau_in) * t *
                 exp(-1/tau_in*t)
④      d/dt V_m == -1/Tau * V_m +
                 1/C_m*I_shape
      end
    else:
      r = r - 1
    end
  end

end
```

Figure 4: Modeling an integrate-and-fire neuron in NESTML. Left panel: using the imperative approach calculating `V_m` explicitly (see ① and ②). Right panel: just specifying the shape of the synaptic current (see ③) and the differntial equation for `V_m` (see ④).

Calculating the matrix for propagating the state is usually a time consuming manual task in NEST. The possibility to write models in a declarative fashion thus considerably reduces the work required to define new models. With the imperative approach still available, we don't have to sacrifice control over other parts of the neuron dynamics, which can nonetheless be expressed as procedural code.

The detailed mathematical and algorithmic techniques for transforming neural dynamics equations to efficient and accurate C++ code are out of scope of the current manuscript, and will be published in a follow-up article.

### 4.4    Implementation of NESTML

NESTML is implemented using the MontiCore [KRV08, KRV07] language workbench, which enables an agile development of DSLs. Based on a context-free grammar, MontiCore defines concrete and abstract representation as abstract syntax tree (AST) and provides infrastructure for checking the compliance to the rules via context conditions [Vö11]. MontiCore supports various mechanisms for heterogeneous language integration, e.g. language aggregation, inheritance, and embedding. These features were used for implementing the NESMTL language inheritance and embedding [Lo13].

The modular design of NESTML gives users the flexibility to exchanges parts of NESTML. For example it enables us to embed Python to be used instead of the Procedural DSL in the future.

MontiCore provides a symbol table infrastructure [Ha15]. The symbol table stores symbols of the model and provides them to the language mechanisms. An example are NESTML components, which provide available functions with their signature, but hide the implementation. NESTML's symbol table automatically handles the resolution of model elements distributed over several files.

All languages of NESTML are strongly typed to allow type compatibility checks within and between models. The checks are performed using information from the symbol table. A constraint inside a model could be one that checks if the `dynamics` block only changes values of the `state` block, while one between models could be a check if a function called from an imported component is actually defined there. The framework to check such context conditions is also provided by MontiCore.

## 5    NESTML Tool Support

We provide a command line interface to the NESTML tools. They process NESTML model descriptions by parsing them, checking context conditions on the described model and generating the C++ model implementation and bootstrapping code for NEST. The code generator is based on the MontiCore generation framework [Sc12], which uses exogenous model-based transformations [MVG06] to integrate the solution code for the differential equation and a template-based system [CH06]. After executing the NESTML tools on a model description, the generated code can be compiled and the model immediately be used in NEST.

During model processing (Figure 5) an abstract syntax tree (AST) is created from the source model and context conditions are checked. From the AST, a SymPy script [Sy14] is generated and executed later by the code generator. For linear neuron models, the script returns the matrix entries of the propagator matrix, or the right hand side of the ODE for use in a solver otherwise. The source AST is transformed by adding these entries as variable declarations to the `internal` block. The altered AST is serialized by pretty printing it again as a NESTML description. This way, the model developer can inspect how the solution of the ODEs is implemented. If the model doesn't contain any differential equations, this step is skipped and code generation is performed on the initial AST.

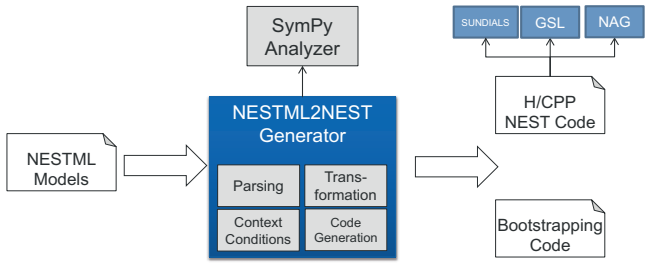

Figure 5: Processing of models in the NESTML frontend. Processing steps include parsing, checking context conditions, model transformations and code generation.

The code generator produces C++ implementation and header files for each model. The integration code consists of a C++ file describing the module and a set of scripts for boot-strapping in NEST. depending on the SymPy analysis, the generated code either contains an explicit implementation of the solution for the ODE, or code that relays the right hand side of the ODE to a numerical solver, e.g. SUNDIALS [Hi05], GSL [Go09] or NAG [HF01].

The frontend is implemented as a regular Java archive available as a download from https://github.com/nest/nestml. It has the following modes of execution:

`parse` Models are parsed and syntactic correctness will be reported.
`contextConditions` Models are parsed and checked against the context conditions. Syntactic and semantic correctness will be reported.
`generate` The code generation workflow will be executed after parsing and checking context conditions.

## 6   Discussion and Outlook

We presented the NEST Modeling Language NESTML to describe spiking neurons and introduced a code generator for the NEST platform. NESTML supports two development paradigms: an imperative scheme based on a procedural language and a declarative scheme using a textual definition of differential equations. Both paradigms can be transparently combined in the same neuron model in order to increase the expressiveness. A dedicated module concept allows a seamless reuse of models and model components. The complexity of model development is decreased by abstracting the implementation and infrastructure details.

Using NESTML, neuron models can be described using domain concepts. The syntax of NESTML is similar to that of the Python programming language, which is well known to the computational neuroscience community. As NESTML is implemented as a MontiCore language, the development of language variants using language inheritance or language embedding is straightforward [Lo13]. This fact can be exploited in the future to develop a family of NESTML languages targeting users with different technical expertise or adding code generation for other simulators.

In order to demonstrate the usefulness of the proposed language, we reformulated about 30% of NEST's models as NESTML in a collaboration with the developers of NEST. The language and tools were generally well received and especially the concise syntax and the code generation pipeline were mentioned as big improvements. NESTML provides a 20 fold reduction of code between model description and generated implementation code. This value includes both the C++ model code as well as the bootstrapping code.

Large-scale modeling of nervous systems requires an abstraction as provided by NESTML to increase modeling capabilities, reusability and maintainability. This is an interesting challenge from the viewpoint of software language engineering and ongoing research will show, how to raise the level of modeling capabilities even further.

NESTML is publically available on https://github.com/nest/nestml. Our ongoing work focuses on the addition of features for the description of synapse models in NESTML. The language and the tools will be evaluated in a more structured way at a community workshop this winter and the feedback will be incorporated into the next public release of NESTML.

**Acknowledgments**   This work is supported by the JARA-HPC Seed Fund *NESTML - A modeling language for spiking neuron and synapse models for NEST*, the *Initiative and Networking Fund* of the Helmholtz Association and the Hemholtz Portfolio Theme *Simulation and Modeling for the Human Brain*. We gratefully acknowledge fruitful discussions with Markus Diesmann and Hans Ekkehard Plesser.

## References

[BG05]   Brette, Romain; Gerstner, Wulfram: Adaptive Exponential Integrate-and-Fire Model as an Effective Description of Neuronal Activity. J Neurophysiol, 94(5):3637–3642, 2005.

[Ch01]   Cheney, J.: Compressing XML with multiplexed hierarchical PPM models.  In: Data Compression Conference, 2001. Proceedings. DCC 2001. pp. 163–172, 2001.

[CH06]   Czarnecki, K.; Helsen, S.: Feature-based Survey of Model Transformation Approaches. IBM Syst. J., 45(3):621–645, July 2006.

[Cr12]   Crook, Sharon; Bednar, James; Berger, Sandra; Cannon, Robert; Davison, Andrew; Djurfeldt, Mikael; Eppler, Jochen; Kriener, Birgit; Furber, Steven; Graham, Bruce; Plesser, Hans Ekkehard; Schwabe, Lars; Smith, Leslie; Steuber, Volker; van Albada, Sacha: Creating, Documenting and Sharing Network Models.  Network: Computation in Neural Systems, 23:131–149, 2012.

[Da13]   Davison, Andrew P.; Diesmann, Markus; Gewaltig, Marc-Oliver; Ghosh, Satrajit S.; Perez, Fernando; Muller, Eilif Benjamin; Bednar, James A.; Thirion, Bertrand; Halchenko, Yaroslav O.: Research topic: Python in Neuroscience II.  Frontiers in Neuroinformatics, 2013. http://journal.frontiersin.org/researchtopic/1591.

[GD07]   Gewaltig, Marc-Oliver; Diesmann, Markus: NEST (NEural Simulation Tool). Scholar-pedia, 2(4), 2007.

[Gl10]   Gleeson, Padraig; Crook, Sharon; Cannon, Robert C.; Hines, Michael L.; Billings, Guy O.; Farinella, Matteo; Morse, Thomas M.; Davison, Andrew P.; Ray, Subhasis; Bhalla, Upinder S.; Barnes, Simon R.; Dimitrova, Yoana D.; Silver, R. Angus: NeuroML: A Language for Describing Data Driven Models of Neurons and Networks with a High Degree of Biological Detail. PLoS Comput Biol, 6(6), 06 2010.

[Go09]   Gough, Brian: GNU scientific library reference manual. Technical report, 2009.

[Go11]   Gorchetchnikov, Anatoli; Raikov, Ivan; Hull, Mike; Le Franc, Yann: Net-work Interchange for Neuroscience Modeling Language (NineML) – Specifica-tion. Technical report, INCF Task Force on Multi-Scale Modeling, July 2011. http://software.incf.org/software/nineml/wiki/nineml-specification/.

[Ha15]   Haber, Arne; Look, Markus; Mir Seyed Nazari, Pedram; Navarro Perez, Antonio; Rumpe, Bernhard; Völkel, Steven; Wortmann, Andreas: Integration of Heterogeneous Modeling Languages via Extensible and Composable Language Components. In (Hammoudi, Sli-mane; Pires, Luis Ferreira; Desfray, Philippe; Filipe, Joaquim Filipe, eds): Proceedings of the 3rd International Conference on Model-Driven Engineering and Software Devel-opment. SciTePress, Angers, Loire Valley, France, pp. 19–31, February 2015.

[HC97]   Hines, M. L.; Carnevale, N. T.: The NEURON Simulation Environment. Neural Compu-tation, 9(6):1179–1209, aug 1997.

[He12]   Helias, Moritz; Kunkel, Susanne; Masumoto, Gen; Igarashi, Jun; Eppler, Jochen Mar-tin; Ishii, Shin; Fukai, Tomoki; Morrison, Abigail; Diesmann, Markus: Supercomputers ready for use as discovery machines for neuroscience. Front. Neuroinform., 6:26, 2012.

[HF01]   Hoffman, Joe D; Frankel, Steven: Numerical methods for engineers and scientists. CRC press, 2001.

[HH52]   Hodgkin, A. L.; Huxley, A. F.: A Quantitative Description of Membrane Current and Its Application to Conduction and Excitation in Nerve. 117:500–544, 1952.

[Hi05]   Hindmarsh, Alan C; Brown, Peter N; Grant, Keith E; Lee, Steven L; Serban, Radu; Shumaker, Dan E; Woodward, Carol S: SUNDIALS: Suite of nonlinear and differen-tial/algebraic equation solvers. ACM Transactions on Mathematical Software (TOMS), 31(3):363–396, 2005.

[Iz03]   Izhikevich, Eugene M et al.: Simple model of spiking neurons. IEEE Transactions on neural networks, 14(6):1569–1572, 2003.

[KRV07]   Krahn, Holger; Rumpe, Bernhard; Völkel, Steven: Integrated Definition of Abstract and Concrete Syntax for Textual Languages. volume 4735 of LNCS, Nashville, TN, USA, pp. 286–300, October 2007.

[KRV08]   Krahn, Holger; Rumpe, Bernhard; Völkel, Steven: Monticore: Modular Development of Textual Domain Specific Languages. volume 11 of LNBIP, Zurich, Switzerland, pp. 297–315, July 2008.

[Ku14]   Kunkel, Susanne; Schmidt, Maximilian; Eppler, Jochen Martin; Plesser, Hans Ekkehard; Masumoto, Gen; Igarashi, Jun; Ishii, Shin; Fukai, Tomoki; Morrison, Abigail; Diesmann, Markus; Helias, Moritz: Spiking network simulation code for petascale computers. Fron-tiers in Neuroinformatics, 8(78), 2014.

[La07]   Lapicque, L.: Recherches quantitatives sur l'excitation electrique des nerfs traitee comme une polarization. J. Physiol. Pathol. Gen, 9:620–635, 1907.

[Lo13]   Look, Markus; Navarro Pérez, Antonio; Ringert, Jan Oliver; Rumpe, Bernhard; Wortmann, Andreas: Black-box Integration of Heterogeneous Modeling Languages for Cyber-Physical Systems. In (Combemale, B.; De Antoni, J.; France, R. B., eds): Proceedings of the 1st Workshop on the Globalization of Modeling Languages (GEMOC). volume 1102 of CEUR Workshop Proceedings, Miami, Florida, USA, 2013.

[MP43]   McCulloch, Warren S.; Pitts, Walter: A logical calculus of the ideas immanent in nervous activity. The bulletin of mathematical biophysics, 5(4):115–133, 1943.

[Mu09]   Muller, Eilif; Bednar, James A.; Diesmann, Markus; Gewaltig, Marc-Oliver; Hines, Michael; Davison, Andrew P.: Research topic: Python in Neuroscience. Frontiers in Neuroinformatics, 2009. http://journal.frontiersin.org/researchtopic/8.

[MVG06]   Mens, Tom; Van Gorp, Pieter: A Taxonomy of Model Transformation. Electron. Notes Theor. Comput. Sci., 152:125–142, March 2006.

[Ni01]   Nicholls, John G.; Martin, Robert A.; Wallace, Bruce G.; Fuchs, Paul A.: From neuron to brain. Sinauer Associates, Sunderland, MA, 4 edition, 2001.

[Ra11]   Raikov, Ivan; Cannon, Robert; Clewley, Robert; Cornelis, Hugo; Davison, Andrew; De Schutter, Erik; Djurfeldt, Mikael; Gleeson, Padraig; Gorchetchnikov, Anatoli; Plesser, Hans Ekkehard; Hill, Sean; Hines, Mike; Kriener, Birgit; Le Franc, Yann; Lo, Chung-Chan; Morrison, Abigail; Muller, Eilif; Ray, Subhasis; Schwabe, Lars; Szatmary, Botond: NineML: the network interchange for neuroscience modeling language. BMC Neuroscience, 12(Suppl 1):P330, 2011.

[RD99]   Rotter, Stefan; Diesmann, Markus: Exact digital simulation of time-invariant linear systems with applications to neuronal modeling. Biological cybernetics, 81(5-6):381–402, 1999.

[Sc12]   Schindler, Martin: Eine Werkzeuginfrastruktur zur agilen Entwicklung mit der UML/P. Aachener Informatik-Berichte, Software Engineering, Band 11. Shaker Verlag, 2012.

[Sh91]   Shepherd, Gordon M.: Foundations of the Neuron Doctrine. Oxford Univ. Press, 1991.

[St14]   Stimberg, Marcel; Goodman, Dan F. M.; Benichoux, Victor; Brette, Romain: Equation-oriented specification of neural models for simulations. Frontiers in Neuroinformatics, 8(6), 2014.

[Sy14]   SymPy Development Team: . SymPy: Python library for symbolic mathematics, 2014. http://www.sympy.org.

[Vö11]   Völkel, Steven: Kompositionale Entwicklung domänenspezifischer Sprachen. Aachener Informatik-Berichte, Software Engineering 9. Shaker Verlag, 2011.

[Ye04]   Yergeau, François; Bray, Tim; Paoli, Jean; Sperberg-McQueen, C Michael; Maler, Eve: Extensible markup language (XML) 1.0. W3C Recommendation, 4, 2004.